\documentclass[12pt]{article}
\usepackage{amsmath,amssymb,graphicx,mathrsfs,hyperref}

\newcommand{\be}{\begin{equation}}
\newcommand{\ee}{\end{equation}}
\newcommand{\bea}{\begin{eqnarray}}
\newcommand{\eea}{\end{eqnarray}}
\newcommand{\bn}{\overline {\nabla}}
\def\({\left(} \def\){\right)}
\begin{document}
\title{\vspace{-1.8in} \begin{flushright} {\footnotesize CERN-PH-TH/2011-139}  \end{flushright}
\vspace{3mm}
\vspace{0.3cm}  Evaluating the Wald Entropy \\ from two-derivative terms in quadratic actions}
\author{\large Ram Brustein${}^{(1,2)}$, Dan Gorbonos${}^{(3)}$,
 Merav Hadad${}^{(1,4)}$ and A.J.M. Medved${}^{(5,6)}$ \\
 \hspace{-.5in} \vbox{
 \begin{flushleft}
  $^{\textrm{
(1)\ Department of Physics, Ben-Gurion University,
    Beer-Sheva 84105, Israel}}$  \\ $^{\textrm{
 (2)  CERN, PH-TH Division, CH-1211, Gen\`eve 23,  Switzerland}}$  \\ $^{\textrm{
(3) Theoretical Physics Institute,
University of Alberta,
Edmonton, Alberta, Canada T6G 2G7 }}$ \\
$^{\textrm{ (4) Department of Natural Sciences, The Open University of Israel,
 P.O.B. 808, Raanana 43107, Israel }}$\\
$^{\textrm{ (5)  Department of Physics and Electronics, Rhodes University,
  Grahamstown 6140, South Africa }}$ \\
$^{\textrm{ (6)  School of Physics, KIAS,
  Dongdaemun-gu, Seoul 130-722,  Korea}}$ \\ \small
    E-mail: ramyb@bgu.ac.il,\ gorbonos@phys.ualberta.ca
meravha@openu.ac.il,\ j.medved@ru.ac.za
\end{flushleft}
}}
\date{}
\maketitle

\begin{abstract}
We evaluate the Wald Noether charge entropy for a black hole in
generalized theories of gravity. Expanding the  Lagrangian to
second order in gravitational perturbations, we show
that contributions to the entropy density originate only from the coefficients  of two-derivative terms.  The same considerations are extended to include matter fields and to show that arbitrary powers of matter fields and their symmetrized covariant derivatives cannot contribute to the entropy density. We also explain how to use the linearized gravitational field equation rather than quadratic actions to obtain the same results. Several explicit examples are presented that allow us to clarify  subtle points in the derivation and application of our method.

\end{abstract}
\newpage

\section{Introduction}

The Bekenstein--Hawking law \cite{bek,hawk},  relates
the entropy $S_{BH}$ to the horizon area $A$ in units of Newton's constant for a black hole in Einstein's theory of gravity,
\be
S_{BH}\;=\;\frac{A}{4 G_N}\;\;.
\ee
This relation suggests that the  entropy $S$ should be  purely geometric,
defined strictly at the black hole
horizon and should satisfy the first law of black hole
mechanics,
\be
T_H dS \; = \; dM \;.
\label{first}
\ee
Here,  $M$
is the conserved or ADM mass of
the black hole  (other conserved charges have been neglected for
simplicity) and $\;T_H=\kappa/2\pi\;$ is the Hawking temperature in terms
of the surface gravity $\kappa$.

The mass $M$ and $\kappa$ are well defined for
a stationary black hole in any theory of gravity, and so their definitions do not need to be modified.
Wald \cite{WALD,IW} proposed  a definition of the entropy
that  fulfills all of the above requirements for general theories of gravity. The Wald entropy $S_W$ has a clear geometric interpretation
through its identification with  the Noether charge for spacetime diffeomorphisms. Further, $S_W$ can always be
cast as a closed integral over a cross-section of
the horizon ${\cal H}$,
\be
S_W=\oint_{\cal H} s_W dA\;,
\ee
with $s_W$ being the entropy per unit of  horizon cross-sectional area.
For a  $D$-dimensional spacetime with metric $\;ds^2 = g_{tt}dt^2
+g_{rr}dr^2+ \sum_{i,j=1}^{D-2}\sigma_{ij}dx_idx_j \;$,
 $\;dA=\sqrt{\sigma}dx_1\dots dx_{D-2}\;$.

The actual Wald formula and how it comes about is briefly reviewed in
Section~2. Meanwhile,
it has since been shown by three of the current authors \cite{BGH} how  Wald's entropy density $s_W$ could also be extracted directly from the gravitational
action. This, through a process of expanding the Lagrangian to quadratic order in the perturbations of the metric and evaluating it at the horizon.  The density $s_W$ can then be identified with the coefficient of the  kinetic terms for the $r,t$-polarized gravitons $h_{rt}$. This coefficient measures the strength of the gravitational coupling for the same
gravitons. The advantage of this method is that it identifies in a straightforward way the correct units in which the area of the horizon should be measured to give the correct value of the entropy. It also provides a way to decide which terms in the expanded action can contribute to the entropy.

Hence, one should always be able to obtain the Wald entropy
by, first,  expanding the Lagrangian around the background solution
and, then, reading off the horizon value of the coefficient of the relevant kinetic terms; for instance, $\nabla^a h_{rt}\nabla_a h_{rt}\;$.
Equivalently and generally easier to implement, one may read  the same coefficients off of terms like $\nabla^a\nabla_a h_{rt}$  in  the linearized field equation.

For two-derivative theories of gravity, for instance, Einstein's and $F({\cal R}$) theories, the procedure is as straightforward as just described. On the other hand, for theories with four or more derivatives, this seemingly simple process can become rather subtle.
Our current motivations are to provide a well-defined prescription
for identifying the kinetic terms and to better understand why
Wald's formulation still works for these higher-derivative cases.

In this paper,  we establish by explicit calculations
that, for a completely generic theory of  gravity,  the  kinetic contributions are indeed the only contributions to the Wald entropy. Further, we verify that the coefficients of these kinetic terms are always sufficient to reproduce Wald's formula for any number of derivatives that may appear explicitly or implicitly in Lagrangian. In the process we clarify the detailed properties of the Wald entropy  that lead to such results.  We then show that adding matter fields
does not alter any of our results.  We identify the correct form of the gravitational field equation that is suitable for
calculations of the entropy via the linearized field equations.

The rest of the paper proceeds as follows.
The next section briefly  reviews the standard  derivation of the Wald formula.
The new material begins in Section~3, where we consider a general theory of gravity and expand the Lagrangian to second order in perturbations.  Then, using basic
properties of the metric, Riemann tensor and horizon generators,
we  distinguish between terms that can possibly contribute to the entropy and  those that  cannot.
We go on to verify that the surviving terms do indeed lead to an entropy in agreement with Wald's expression.
In Section~4, the previous considerations are then  extended to
generic theories of both gravity and  matter.  Our attention turns,
in Section~5, to
the gravitational field equation for a generalized theory.
In Section~6, some of specific models  are used to
illustrate our earlier analysis. Section~7 contains a summary and some concluding comments.

\section{A brief review of the Wald Noether charge entropy}

In this section, we recall the derivation of the Wald entropy,
following  Jacobson, Kang and Myers
\cite{JKM}. Our goal is to make the paper self-contained. For brevity, we skip over  the many subtleties and caveats in the derivations.

One starts  by varying  a given Lagrangian density $L$ with respect to all the fields $\{\psi\}$, including the metric. In condensed notation (with all tensor indices suppressed),
\be
\delta L\;=\; {\cal E}\cdot \delta \psi \;+\; d\theta\left(\delta \psi
\right)\;,
\label{jkm1}
\ee
where $\;{\cal E}=0\;$ are the equations of motion and the dot represents a summation over all fields and  contractions of tensor indices. Also, $d$ denotes a total derivative, so that $\theta$ is a boundary term.

Let  $\pounds_{\xi}$ be a Lie derivative acting along some vector
field $\xi$.
Then, given the diffeomorphism invariance of the theory,
$\;\delta_{\xi}\psi=\pounds_{\xi}\psi\;$
and $\;\delta_{\xi}L=\pounds_{\xi} L=d\left(\xi\cdot L\right)\;$.
These and  Eq.~(\ref{jkm1}) can be used to identify the
associated Noether  current $J_{\xi}$,
\be
J_{\xi}\;=\; \theta \left(\pounds_{\xi}\psi\right) -\xi\cdot{L}\;.
\label{jkm2}
\ee
The point being that $\;dJ_{\xi}=0\;$ when $\;{\cal E}=0\;$, and so
there must be an associated ``potential'' $Q_{\xi}$ such that
$\;J_{\xi}=dQ_{\xi}\;$.
Now, if $D$ is the dimension of the spacetime and
${\cal S}$ is a $D-1$  hypersurface with a
$D-2$ spacelike boundary $\partial{\cal S}$, then
\be
\int_{{\cal S}} J_{\xi} \;=\;\int_{\partial{\cal S}} Q_{\xi}
\ee
is the associated Noether charge.

Wald showed \cite{WALD} and later proved rigorously
\cite{IW} that the black hole first law (\ref{first})
is satisfied when the entropy is defined
in terms of a specific Noether charge. Choosing  the surface ${\cal }S$ as the
horizon
${\cal H}$ and the vector field $\xi$ as the  horizon Killing
vector $\chi$ (with
its surface gravity normalized to unity~\footnote{This particular normalization
for the Killing vector will be assumed throughout the paper.}), Wald identified
the entropy as
\be
S_W\; \equiv\; 2\pi \oint_{\cal H} Q_{\chi}\;.
\label{jkm3}
\ee
Since $\;\chi=0\;$
on the horizon,~\footnote{More accurately, at the bifurcation surface
of the horizon. This is one of the many caveats that are dealt with
in \cite{JKM}.}
the right-most term in Eq.~(\ref{jkm2}) does not contribute
to $S_W$.

One of the main advantages of Wald's formula is the simplicity of
$Q$. To understand this, let
us start with the Killing identity
\cite{TEXT}
\be
\nabla_c \nabla_a \chi_b \;=\; -{\cal R}_{abcd}\chi^d \;,
\label{kill}
\ee
where ${\cal R}_{abcd}$ is the Riemann tensor.
After repeated applications of this relation,
the most general form of the integrand in
Eq.~(\ref{jkm3}) can be expressed as
\be
Q^{ab}\epsilon_{ab} \;=\;
\left[{\cal B}^{ab}_{\;\;\;\;\;c} \chi^c
+{\cal C}^{ab}_{\;\;\;\;\;\;cd}\nabla^c\chi^d\right]
\epsilon_{ab} \;,
\ee
where ${\cal B}_{abc}$ and ${\cal C}_{abcd}$ are theory-dependent background tensors,
while $\;\epsilon_{ab} \equiv \nabla_a\chi_b\;$ is the binormal vector
for the horizon.  For future reference,
\be
\epsilon_{ab}\;=\;-\epsilon_{ba}
\label{anti}
\ee

From the  definition of $\epsilon_{ab}$ and since $\;\chi^a = 0\;$ (on ${\cal H}$),
it follows that the integrand simplifies to
\be
Q^{ab}\epsilon_{ab} = {\cal C}^{abcd}\epsilon_{ab}
\epsilon_{cd} \;
\ee
or
\be
S_W \;=\; 2\pi \oint_{{\cal H}}{\cal C}^{abcd}\epsilon_{ab}\epsilon_{cd}dA\;.
\label{jkm9}
\ee

We continue to sketch the
analysis of  \cite{JKM}
for theories with a Lagrangian
$\;{\cal L}= {\cal L}[g_{ab},{\cal R}_{abcd}]\;$.

First, the variation of the density  $\;L=\sqrt{-g}{\cal L}\;$ is found
to yield
\be
\delta L\;=\;-2\nabla_a\left({\cal X}^{abcd}\nabla_c \delta g_{bd}\sqrt{-g}
\right)\;+\;\cdots\;,
\label{jkm5}
\ee
where dots are meant as terms that end up being irrelevant to
the   Wald entropy
and
\be
{\cal X}^{abcd}\;\equiv\frac{\partial\cal L}{\partial {\cal R}_{abcd}}\;.
\label{calx}
\ee
It follows from Eq.~(\ref{jkm1}) that  Eq.~(\ref{jkm5}) leads to
\be
\theta \;=\; -2n_a{\cal X}^{abcd}\nabla_c \delta g_{bd}\sqrt{\gamma}\;+\;\cdots\;,
\ee
where $n^a$ is the unit normal vector and  $\gamma_{ab}$ is the induced metric
for the chosen surface
${\cal S}$.

For an arbitrary diffeomorphism
$\;\delta g_{ab} \;=\; \nabla_a \xi_b + \nabla_b\xi_a\;$, the associated Noether
current is then equal to
\be
J\;=\;-2\nabla_a\left({\cal X}^{abcd}\nabla_c\left(\nabla_b\xi_a +
\nabla_a\xi_b\right)n_a \sqrt{h}
\right)\;+\;\cdots\;.
\label{jkm6}
\ee

Let us now specialize to the horizon $\;{\cal S}\to{\cal H}\;$
and (normalized) Killing vector $\;\xi^a\to\chi^a\;$; so that
$\;\ n_a\sqrt{h}\to\epsilon_a\sqrt{\sigma}\;$ with $\;\epsilon_a\equiv
\epsilon_{ab}\chi^b\;$.
Then, using the symmetries of ${\cal X}^{abcd}$ (inherited from ${\cal R}^{abcd}$)
along with  Eq.~(\ref{anti}),
one can eventually translate Eq.~(\ref{jkm6}) into
\be
J\;=\; -2\nabla_b\left({\cal X}^{abcd}
\nabla_c\chi_d \epsilon_a\sqrt{\sigma} \right)\;+\;\cdots\;.
\ee
In this form, the potential is
\be
Q\;=\;  -{\cal X}^{abcd}
\epsilon_{ab}\epsilon_{cd}\sqrt{\sigma} \;+\;\cdots\;,
\ee
and so
\be
S_W\;=\;  -2\pi\oint_{\cal H} {\cal X}^{abcd}
\epsilon_{ab}\epsilon_{cd} dA\;,
\label{skip}
\ee
with ${\cal X}^{abcd}$ defined in Eq.~(\ref{calx}).

\section{Evaluating the Wald entropy}

The main goal  of this section is to establish our claims that only
the kinetic terms for the $h_{rt}$ gravitons can
contribute to the Wald formula and that the entropy can be deduced simply by reading off their coefficients. We accomplish this for a generic theory by, first, identifying all possible kinetic terms
in the quadratic expansion of the  Lagrangian and, then,
verifying that the coefficients of the relevant terms
produce a result that agrees with  Wald's expression.  Along the way, we find that the contributions to the Wald entropy arise from a specific class of terms in the quadratically expanded Lagrangian, those terms that contain a second-order expansion of the Riemann tensor.

\subsection{Preliminaries}

We will begin with a pure gravitational theory.  The Lagrangian for such a theory can be expressed  in terms of the metric, the Riemann tensor and its symmetrized covariant derivatives,
\be
{\cal L}\;=\;{\cal L}\left[g_{ab},{\cal R}_{abcd},
\nabla_{a_1}{\cal R}_{abcd},
\nabla_{\left(a_1\right. }\nabla_{\left. a_2\right)}{\cal R}_{abcd},\ldots\right]\;,
\ee
where the ellipsis denote increasing numbers of symmetrized covariant
derivatives acting on the Riemann tensor. The derivatives can be  expressed in such a symmetrized form,  as any anti-symmetric combination can be converted into a Riemann tensor.

Our objective is to expand the Lagrangian density
$\sqrt{-g}{\cal L}$ to second order in the  metric
perturbations, $\;h_{ab}=g_{ab} - g^{(0)}_{ab}\;$, and then isolate
the two-derivative terms which we call ``kinetic terms". We should also consider terms that have four or more
derivatives, as clarified below.
We will show that any  kinetic term on the horizon can be expressed as
\be
\left[{\cal A}^{abcd}\right]^{(0)}\bn^e h_{ab}\bn_e h_{cd}\;.
\label{kinetic}
\ee
Here and in what follows,  ${\cal A}^{a_1 a_2\dots}$ represents an
arbitrary tensor built out of the Riemann tensor, its symmetrized
derivatives and the metric.
A numeric superscript on a tensor
denotes its order in $h$'s and $\bn$  is a zeroth-order covariant derivative.

We can prove that a kinetic term is always of the form (\ref{kinetic}), which is a four-index tensor with the two derivatives contracted with each other.  Let us begin with the most general  term carrying exactly
 two gravitons and exactly two derivatives.
Repeatedly integrating by parts and discarding surface terms and ``mass terms" which have no derivatives acting on a graviton,
we eventually arrive at
\be
\left[\widetilde{\cal A}^{abcdef}\right]^{(0)}\bn_a h_{bc} \bn_d h_{ef}\;.
\ee
Here, the background tensor contains no explicit derivatives
and is denoted by  $\widetilde{\cal A}$.

Given our Lagrangian, the background tensor  $\left[\widetilde{\cal A}^{abcdef}\right]^{(0)}$  is built out
of the tensors  $\left[g^{ab}\right]^{(0)}$  and
\be
\left[R^{abcd}\right]^{(0)}\;\propto\; \left[g^{ac}g^{bd}-g^{ad}g^{bc}\right]^{(0)}
\;.
\label{bimet}
\ee
The last expression follows  from
the fact that, when evaluated on a stationary horizon, any symmetric  tensor
${\cal A}_{sym}^{ab}$ can be expressed
as (see Section 5 of \cite{MMV})
\be
{\cal A}^{ab}_{sym} \;=\; {\cal A} g^{ab}\;,
\ee
for some scalar ${\cal A}$.~\footnote{This statement is valid on the horizon's bifurcation surface. However,
as  Wald's integral expression can be evaluated
over an arbitrarily  chosen cross-sectional slice   \cite{JKM}, one can always calculate on the  bifurcating slice without  loss of generality.}
The background Ricci tensor $\left[{\cal R}^{ab}\right]^{(0)}$ is of
this form. Then, given that
$\;{\cal R}^{ac}= g_{bd}{\cal R}^{abcd}\;$
and $\;{\cal R}^{ad}=-g_{bc}{\cal R}^{abcd}\;$,
the form of Eq.~(\ref{bimet}) follows.

And so we have found that any  index on  $\left[\widetilde{\cal A}^{abcdef}\right]^{(0)}$
is associated with  a metric tensor.
Then, any index on
\be
\bn_a h_{bc} \bn_d h_{ef}
\ee
must be contracted   with  one of the
other five  indices.

Let us now choose the transverse and traceless
gauge for the gravitons,~\footnote{This choice is for convenience only, as the
Wald entropy is gauge invariant.}
\be
\bn_a h^a_{\ b},\ h^a_{\ a} \;=\;0\;.
\label{gauge}
\ee
So that contractions such as
\be
\left[\widetilde{\cal A}^{abcd}\right]^{(0)}\bn_e h^e_{\ a} \bn_b h_{cd} \;
\label{above}
\ee
and
\be
\left[\widetilde{\cal A}^{abcd}\right]^{(0)}\bn_e h_{ab} \bn_c h^e_{\ d}
\label{above5}
\ee
vanish.

To show that the term (\ref{above5}) vanishes, we first argue that it must be of the form
\be
\left[\widetilde{\cal A}^{bd}\right]^{(0)}\bn_e h^f_{\ b} \bn_f h^e_{\ d}\;,
\ee
otherwise a trace $h^a_{\ a}$ appears.
Integrating by parts with $\bn_e$, we have
\be
\left[\widetilde{\cal A}^{bd}\right]^{(0)}\bn_e h^f_{\ b} \bn_f h^e_{\ d}\;
=\;-\bn_e\left(\left[\widetilde{\cal A}^{bd}\right]^{(0)}\right)h^f_{\ b}\bn_f h^e_{\ d}
\;-\; \left[\widetilde{\cal A}^{bd}\right]^{(0)}h^f_{b}\bn_e\bn_f h^e_{\ d} \;.
\label{above2}
\ee
All derivatives can be eliminated from the second term:
\bea
\bn_e\bn_f h^e_{\ d} \;&=&\; \bn_f\bn_e h^e_{\ d} \;+\; \left[\bn_e,\bn_f\right]h^e_{\ d}
\nonumber \\
\;&=& \left[{\cal R}_{ef\;\;\;\;\;a}^{\;\;\;\;\;e}\right]^{(0)}h^a_{\ d}
\;+\;\left[{\cal R}_{efd}^{\;\;\;\;\;\;a}\right]^{(0)}h^e_{\ a}\;,
\eea
which follows from Eq.~(\ref{gauge}) and  the standard identities relating commutators of derivatives to the Riemann tensor.

Integrating  by parts $\bn_f$ in the first term in
Eq.~(\ref{above2}), we find
\be
-\bn_e\left(\left[\widetilde{\cal A}^{bd}\right]^{(0)}\right)h^f_{\ b}\bn_f h^e_{\ d}=\bn_f\bn_e\left(\left[\widetilde{\cal A}^{bd}\right]^{(0)}\right)h^f_{\ b} h^e_{\ d}
\;+\;
\bn_e\left(\left[\widetilde{\cal A}^{bd}\right]^{(0)}\right)\left(\bn_f h^f_{\ b}\right) h^e_{\ d}
\;,
\ee
so
\be
-\bn_e\left(\left[\widetilde{\cal A}^{bd}\right]^{(0)}\right)h^f_{\ b}\bn_f h^e_{\ d}=\left[{\cal B}_{fe}^{\;\;\;\;\;bd}\right]^{(0)}h^f_{\ b} h^e_{\ d}
\;;
\ee
where, besides transversality,  we have redefined the background tensor
$\;\left[{\cal B}^{febd}\right]^{(0)}\equiv\bn^f\bn^e\left[\widetilde{\cal A}^{bd}\right]^{(0)}\;$
to emphasize that the above is really a mass term.  Hence the term  (\ref{above5}) is a mass term.

One can  now see that the two $\bn$'s in a kinetic term must  be
contracted, leading to the claimed form in Eq.~(\ref{kinetic}).
But what about $h^2$  terms with greater numbers of derivatives?
Terms having more than two $\bn$'s can also make
kinetic contributions, as follows.

Derivatives have to be contracted in pairs since they cannot be contracted with a graviton index because of the gauge condition.
Consequently, given a generic term with exactly $2n$ derivatives
\be
\left[\widetilde{\cal B}^{a_1a_2\dots a_{2n-1} a_{2n}}\right]^{(0)}
\bn_{a_1}\bn_{a_2}\dots\bn_{a_{2n-1}}\bn_{a_{2n}}
\left(\left[\widetilde{\cal A}^{abcd}\right]^{(0)}h_{ab}h_{cd}\right)\;,
\ee
one can use integration by parts,
commutator relations like
\be
\left[\bn_a, \bn_b\right] h_{cd} \;=\;
\left[{\cal R}_{abc}^{\;\;\;\;\;\;e}\right]^{(0)}
h_{ed} + \left[{\cal R}_{abd}^{\;\;\;\;\;\;e}\right]^{(0)}
h_{ce}\;,
\ee
symmetry properties of the background  such as
\bea
\left[{\cal R}^{abcd}\right]^{(0)} \;&=&\;
-\left[{\cal R}^{bacd}\right]^{(0)} \;=\;
-\left[{\cal R}^{abdc}\right]^{(0)} \;=\;
\left[{\cal R}^{dcba}\right]^{(0)} \\
\;&=&\;-\left[{\cal R}^{cabd}\right]^{(0)}\;-\;\left[{\cal R}^{bcad}\right]^{(0)}
\;\label{jac}
\eea
and the  gauge conditions (\ref{gauge}), to manipulate the expression into a power series in $\Box\equiv\bn_e\bn^e$ acting on the gravitons. (Similar to our manipulation of Eq.~(\ref{above5}) into a series terminating at $\Box^0$.) That is,
\be
\sum_{j=0}^{n}\left[{\cal A}^{abcd}\right]_j^{(0)} h_{ab}\Box^j h_{cd}\;.
\ee

For instance, a term initially containing  four derivatives
and two gravitons  can, after sufficient manipulations, be
reduced to  a combination of  three  forms:
\be
\left[\widetilde{\cal A}^{abcd}\right]_{2}^{(0)} h_{ab}\Box^2 h_{cd}\;,
\left[{\cal A}^{abcd}\right]_{1}^{(0)} h_{ab} \Box h_{cd}\;,
\left[{\cal A}^{abcd}\right]_{0}^{(0)} h_{ab}h_{cd}\;.
\ee
This outcome  follows from our earlier discussion, where it was shown that the background geometry can only act on gravitons
so as to contract  indices  {\em or} yield pairs of  contracted derivatives. The essential point
is that higher-derivative terms can contribute to the kinetic terms and therefore to the Wald entropy, provided that
all but two of  the derivatives act  on the background tensors.

\subsection{Expanding the Lagrangian}

Let us now begin the formal calculation by writing down the second-order
expansion of  $\sqrt{-g}{\cal L}$:
\be
\delta{\hat{\cal L}}^{(2)}\;=\; \left[\frac{\partial\left(\sqrt{-g}\cal L\right)}
{\partial g_{ab}}
\frac{\delta g_{ab}}{\sqrt{-g}}\right]^{(2)}
 +  \left[\frac{\partial{\cal L}}
{\partial {\cal R}_{abcd}}\delta {\cal R}_{abcd}\right]^{(2)}
+\left[\frac{\partial{\cal L}}{\partial[\nabla_{a_1} {\cal R}_{abcd}]}
\delta\nabla_{a_1}{\cal R}_{abcd}\right]^{(2)}
\nonumber \ee
\be
\hspace{5cm}+\left[\frac{\partial{\cal L}}
{\partial[\nabla_{\left(a_1\right.} \nabla_{\left. a_2\right)}{\cal R}_{abcd}]}
\delta\nabla_{\left(a_1\right. }\nabla_{\left. a_2\right)}{\cal R}_{abcd}\right]^{(2)}
+\ldots\;,
\label{exp} \ee
where
\be
\delta{\hat{\cal L}}\;\equiv\; \frac{\delta\left[\sqrt{-g}{\cal L}\right]}
{\sqrt{-g}}\;
\ee
and the ellipsis has  now be used to denote variations
with respect to ever-increasing numbers of symmetrized derivatives.

To proceed, we  follow
an  iterative procedure
that was first laid out in the third section
of \cite{IW}.  The basic  idea is that a term like
\be
\left[\frac{\partial{\cal L}}
{\partial[\nabla_{\left(a_1\right.}\ldots \nabla_{\left. a_j\right)}{\cal R}_{abcd}]}
\delta\nabla_{\left(a_1\right.}\ldots\nabla_{\left. a_j\right)}{\cal R}_{abcd}
\right]^{(2)}
\ee
can always be re-expressed as
\be
\left[\frac{\partial{\cal L}}
{\partial[\nabla_{\left(a_1\right.}\ldots \nabla_{\left. a_j\right)}{\cal R}_{abcd}]}
\nabla_{\left(a_1\right.}
\delta\nabla_{a_2 }\ldots\nabla_{\left. a_j\right)}{\cal R}_{abcd}\right]^{(2)}
\ee
plus terms that are proportional to
$\nabla_{a_1}\delta g $.
Then, integrating everything by parts, one has (up to surface terms)
\be
-\left[\nabla_{\left(a_1\right.}\left(\frac{\partial{\cal L}}
{\partial[\nabla_{\left(a_1\right.}\ldots \nabla_{\left. a_j\right)}{\cal R}_{abcd}]}\right)
\delta\nabla_{a_2}\ldots\nabla_{\left. a_j\right)}{\cal R}_{abcd}\right]^{(2)}
\ee
plus terms that are proportional to $\delta g$.
One can repeat this process $j-1$ more times until
obtaining
\be
(-)^{j}\left[\nabla_{\left(a_j\right.}\ldots \nabla_{\left.a_1\right)}
\left(\frac{\partial{\cal L}}
{\partial[\nabla_{\left(a_1\right.}\ldots \nabla_{\left. a_j\right)}{\cal R}_{abcd}]}\right)
\delta{\cal R}_{abcd}\right]^{(2)} \;,
\ee
along with  a collection of terms that are proportional to
$\delta g$ (as well as surface terms).

Consequently, we can reorganize the expansion (\ref{exp}) as follows
(up to surface terms):
\be
\delta{\hat{\cal L}}^{(2)}\;=\; \left[{\cal W}^{ab}\delta g_{ab}\right]^{(2)}
\;+ \; \left[{\cal X}^{abcd}\delta{\cal R}_{abcd}\right]^{(2)}\;,
\label{W1}
\ee
where ${\cal W}^{ab}$ is some tensorial function of the geometry
(with its precise form
being irrelevant to what follows~\footnote{But note that, for   $\;{\cal L}={\cal L}[g_{ab},{\cal R}_{abcd}]\;$ theories,
$\;{\cal W}^{ab}=\partial {\cal L}/\partial g_{ab} +
\frac{1}{2}g^{ab}{\cal L} \;$.}) and
\bea
{\cal X}^{abcd}\;\equiv\; \frac{\partial{\cal L}}{\partial {\cal R}_{abcd}}
 \;&-&\;\nabla_{a_1}\left(\frac{\partial{\cal L}}
{\partial[\nabla_{a_1} {\cal R}_{abcd}]}\right) \nonumber \\
\;&+&\;\nabla_{\left(a_1\right. }\nabla_{\left. a_2\right)}
\left(\frac{\partial{\cal L}}
{\partial[\nabla_{\left(a_1\right.} \nabla_{\left. a_2\right)}{\cal R}_{abcd}]}\right)
\;+\; \ldots\;
\label{W2}
\eea
is a tensor with the same  symmetry properties of the
Riemann tensor. Notice that ${\cal X}^{abcd}$ is the generalized version of the same-named tensor in Eq.~(\ref{calx}).

\subsection{Isolating the kinetic terms}

Now, since    $\;\delta g_{ab} =g_{ab}-g^{(0)}_{ab}\;$, it follows
that the complete non-expanded
form of the ${\cal W}$ term in Eq.~(\ref{W1})
contains  a factor   $g_{ab}$.
Such a term can be  dismissed,
as integration by parts can be used to kill
off any would-be kinetic
contribution.~\footnote{Let us emphasize that this is different than
starting with, say,
$\;\delta {\cal R}^{(2)} \sim \bn h\bn h+\cdots\;$
and then integrating by parts to come up with $\; h\bn\bn h\;$.
In this example, there may be an undifferentiated  $h$,
but it did not originate from a metric.}

To better understand this argument, suppose that we
had the generic form
\be
\left[\nabla_{\left(a_1\right.}\cdots\nabla_{\left. a_j\right)} h_{ab}\right]
{\cal A}^{abcd;a_1\ldots a_j} (g_{cd}^{(0)}+h_{cd})\;.
\ee
Here, all expressions are regarded as
full unexpanded expressions except
where indicated.
To create a kinetic term,
it is then necessary to move derivatives (via integration by parts)
until obtaining a combination of the form
\be
  \nabla_{a_j} h_{ab}
\left[\nabla_{a_{j-1}}\cdots \nabla_{a_2}{\cal A}^{abcd;a_1\ldots a_j}\right]
\nabla_{a_1}(g_{cd}^{(0)}+h_{cd})\;.
\ee
But such a term contains the vanishing factor $\;\nabla_{a_1} g_{cd}=0\;$, and so the would-be kinetic term never has a chance to materialize.

Having dismissed the metric variation, let us next focus on the  $\delta {\cal R}_{abcd}$ contribution.
Schematically, this goes as $\;\delta {\cal R} =\nabla \delta \Gamma + \delta \Gamma \delta \Gamma\;$,
with $\;\delta \Gamma = \nabla h\;$.  More precisely,
\be
\delta \left[\Gamma^a_{bc}\right]^{(1)}
\;=\;\frac{1}{2}\left[\bn_b h^a_{\ c} + \bn_ch^a_{\ b} - \bn^ah_{bc}\right]\;,
\ee
from which one obtains
\be
\delta {\cal R}_{abcd}^{(1)}[h] \;=\;
\frac{1}{2} \left[\bn_c\bn_b h_{ad}
\;+\;  \bn_d\bn_a h_{bc}
\;-\; \bn_d\bn_b h_{ac}
\;-\; \bn_c\bn_a h_{bd} \right]\;
\label{expan}
\ee
and
\be
\delta {\cal R}_{abcd}^{(2)}\;=\; \frac{1}{4}
\left[\bn_c h_{ea} \bn_d h^e_{\ b} + \bn_c h_{ea} \bn_b h^e_{\ d}
-\bn_e h_{ca} \bn^e h_{db}
-\bn_a h_{ce} \bn_d h^e_{\ b} \right.
\nonumber \\
\ee
\be
\left. -\bn_a h_{ce} \bn_b h^e_{\ d}
-\bn_c h_{ea} \bn^e h_{db}
+\bn_e h_{ca} \bn_d h^e_{\ b}
+\bn_e h_{ca} \bn_b h^e_{\ d} +\bn_a h_{ce} \bn^e h_{db}\right]
\nonumber \\
\ee
\be
-\; \left\{c \longleftrightarrow d\right\} \;.
\label{bigmess}
\ee

From these expansions, it can be deduced that
\bea
\delta{\hat{\cal L}}_k^{(2)}\;&=&\;
                         \left[{\cal X}^{abcd}\right]^{(0)}
\delta{\cal R}_{abcd}^{(2)} \;+\;  \left[{\cal X}^{abcd}\right]^{(1)}
\delta{\cal R}_{abcd}^{(1)} \\
\;&=&\;
                        \frac{1}{4} \left[{\cal X}^{abcd}\right]^{(0)}
\left(\bn_c h_{ea}\bn_d h_{\ b}^e +\cdots\right) \;+\;
\frac{1}{2}\left[{\cal X}^{abcd}\right]^{(1)}\left( \bn_c\bn_b h_{ad}
+\cdots\right) \;, \nonumber
\label{L-k}
\eea
where the subscript $k$ indicates that we only intend
to retain the kinetic contributions and the ellipses denote the
various permutations of the displayed indices.
The first term on the right-hand side can clearly be kinetic
and is what would normally be attributed
to the Wald entropy. The question then is what becomes of the second term?

To make sense of the second term, it is necessary to expand  the
tensor ${\cal X}^{abcd}$ to first
order. Following the same iterative procedure  as before, we have
\be
\left[{\cal X}^{abcd}\right]^{(1)}
\;=\; \left[{\cal Z}^{abcd;ef}\right]^{(0)}\delta g_{ef}^{(1)}
\;+\;\left[{\cal Y}^{\substack{abcd \\ pqrs}}\right]^{(0)}
\delta{\cal R}_{pqrs}^{(1)}\;,
\label{X-of-h}
\ee
where ${\cal Z}^{abcd;ef}$ is a tensor akin to ${\cal W}^{ab}$ and
\bea
{\cal Y}^{\substack{abcd \\ pqrs}}\;\equiv\;
\frac{\partial{\cal X}^{abcd}}{\partial {\cal R}_{pqrs}}
\;&-&\;\nabla_{a_1}\left(\frac{\partial{\cal X}^{abcd}}
{\partial[\nabla_{a_1} {\cal R}_{pqrs}]}\right)\nonumber\\
\;&+&\;\nabla_{\left(a_1\right. }\nabla_{\left. a_2\right)}
\left(\frac{\partial{\cal X}^{abcd}}
{\partial[\nabla_{\left(a_1\right.} \nabla_{\left. a_2\right)}{\cal R}_{pqrs}]}\right)
\;+\; \ldots\;
\eea
is  a tensor that is  ``two-fold'' Riemann symmetric.

Since $\;\delta g_{ef}=g_{ef}-g^{(0)}_{ef}\;$, we can, as previously discussed,
neglect the first term on the right-hand side
of Eq.~(\ref{X-of-h}). Also
recalling Eq.~(\ref{expan}) for $\delta{\cal R}^{(1)}$, we have
\be
\left[{\cal X}^{abcd}\right]^{(1)}
\;=\;
\frac{1}{2}\left[{\cal Y}^{\substack{abcd \\ pqrs}}\right]^{(0)}
\left[\bn_r\bn_q h_{ps}
+  \bn_s\bn_p h_{qr}
- \bn_s\bn_q h_{pr}
- \bn_r\bn_p h_{qs} \right]\;.
\label{chiyi}
\ee
The insertion of  Eq.~(\ref{chiyi})
into Eq.~(\ref{L-k}) then yields
\bea
\delta{\hat{\cal L}}_k^{(2)}\;&=&\;
 \frac{1}{4}\left[{\cal X}^{abcd}\right]^{(0)}\left( \bn_c h_{ea}\bn_d h_b^{e}
+\cdots\right) \nonumber \\
\;&+&\;\;\;
\frac{1}{4}\left[{\cal Y}^{\substack{abcd \\ pqrs}}\right]^{(0)}
\left(\bn_c  \bn_b h_{ad}\; \bn_r \bn_q h_{ps}+\cdots\right) \;.
\label{L-k2}
\eea
As already mentioned, a four-$\bn$ term might still make a kinetic contribution, so that the second
term can not be dismissed. Nevertheless, we will now proceed to demonstrate that such a term can not contribute to the Wald entropy, simply because this is actually a mass term.

In the Wald-entropy prescription \cite{WALD,IW,JKM}
the relevant  metric perturbations are from the restricted class
$\;\left. h_{ab}\right|_{{\cal H}\;;\;\left\{a,b\right\}=\left\{r,t\right\}}\; $ and
\be
h_{ab}\;=\;\nabla_a \chi_b +\nabla_b \chi_a\;,
\label{pert}
\ee
where $\chi_a$ is a vector field that (at least) limits to the Killing
vector on the horizon.
Actually, as mentioned in Section 1, the Killing vector has been
normalized such that
\be
\epsilon_{ab} \;=\; \nabla_a\chi_b \;,
\label{afterpert}
\ee
 is the horizon binormal, for which
\be
\epsilon_{ab} \;\neq\; 0 \;\;\; {\rm iff} \;\;\;\left\{a\neq b\right\}\;=\;\left\{r,t\right\}\;;
\label{bino}
\ee
and so only the off-diagonal elements $h_{rt}$
are relevant.

Let us now consider a  term such as
$\;\nabla_c\nabla_b \nabla_a \chi_d\;$.
One can use the Killing identity (Eq.~(\ref{kill})) to rewrite this as
\bea
\nabla_c\nabla_b \nabla_a \chi_d \;&=&\;
\nabla_c\left[{\cal R}_{dabe}\chi^e\right]\nonumber\\
\;&=&\;{\cal R}_{dabe}\nabla_c\chi^e\;,
\eea
where the second line is a consequence of  $\;\chi^a=0\;$ on the horizon.
This outcome leads  us to, for instance,
\bea
\bn_c\bn_b h_{ad}\;&=&\; {\cal R}^{(0)}_{dabe}\bn_c\chi^e
+{\cal R}^{(0)}_{adbe}\bn_c\chi^e \nonumber \\
\;&=&\;
{\cal R}^{(0)}_{dabe}\left[\bn_c\chi^e + \bn^e\chi_c\right]
\nonumber \\ \;&=&\;
{\cal R}^{(0)}_{dabe}h_c^e\;,
\eea
where  Eq.~(\ref{pert}) has been used twice and
 the second line comes about from the anti-symmetries of
both the Riemann tensor and the horizon binormal vector;
{\it cf}, Eq.~(\ref{anti}).

It now follows that  the term proportional to ${\cal Y}$ in Eq.~(\ref{L-k2})  can equivalently be written as
\be
\frac{1}{4}\left[{\cal Y}^{\substack{abcd \\ pqrs}}\right]^{(0)}
\left({\cal R}^{(0)}_{dabe} h_c^e\; {\cal R}^{(0)}_{spqw} h_r^w\;+\;\cdots\right) \;;
\ee
so that any  such term is a mass term
as promised.~\footnote{If
the purpose is to calculate the Wald entropy, then
one   cannot ``cheat'' by  having the Killing identity preceded
by an integration by parts.
That is, one is not permitted to turn $\;\delta{\cal R}^{(2)}\sim
\nabla h\nabla h\;$ into  $\;h\nabla\nabla h\sim
h{\cal R}h\;$ and then argue that this is a mass term.}

Hence, the kinetic contribution and the Wald entropy is determined
strictly by  the first part of Eq.~(\ref{L-k2}). Following  the result from Eq.~(23) of \cite{BGH}
(which follows from Eq.~(\ref{bigmess}) and  the symmetry properties
of $\left[{\cal X}^{abcd}\right]^{(0)}$)
\be
\left[{\cal X}^{abcd}\right]^{(0)}\delta {\cal R}^{(2)}_{abcd} \;=\;
\frac{1}{2}\left[{\cal X}^{abcd}\right]^{(0)}\left(\bn^e h_{bc} \bn_e h_{ad}
+ 2 \bn^e h_{ac}\bn_b h_{de}\right)\;
\label{rtoo}
\ee
and recognizing that the second term  can be gauged away,
we  arrive at
\be
\delta{\hat{\cal L}}_k^{(2)}\;=\;
 \frac{1}{2}\left[{\cal X}^{abcd}\right]^{(0)}\;\bn^e h_{bc}\bn_e
h_{ad}\;.
\label{kin-fin}
\ee

We can be even more precise  by remembering that the relevant perturbations
are sourced strictly by the off-diagonal elements in the
$\left\{r,t \right\}$ sector of gravitons. Hence,
\be
\delta{\hat{\cal L}}_k^{(2)}\;=\;
\frac{1}{2}\left[{\cal X}^{rt}_{\;\;\;\;\;rt}\right]^{(0)}
\;\sum_{a\neq b}^{r,t}\bn^e h_{ab}\bn_e
h^{ab}
\;.
\label{kin-fin2}
\ee

\subsection{Interpretations}

Our claim is  that all information about the Wald entropy
is (up to normalization~\footnote{The correct normalization can
always be uniquely fixed by the Einstein Lagrangian.
In this case,
 $\;{\cal X}^{ab}_{\;\;\;\;\;cd}=\frac {1}{2}\left[g^a_cg^b_d-g^a_dg^b_c\right]\;$,
and so  $\;\left[{\cal X}^{rt}_{\;\;\;\;\;rt}\right]^{(0)}= +\frac{1}{2}\;$.})
encoded in the horizon value of the single tensorial component
$\left[{\cal X}^{rt}_{\;\;\;\;\;rt}\right]^{(0)}$,
\be
s_W \;=\; {\cal C}\left[{\cal X}^{abcd}\right]^{(0)}\epsilon_{ab}\epsilon_{cd}\;,
\label{kin-fin3}
\ee
where $s_W$ is the ``Wald entropy density''
(or entropy per unit of  horizon cross-sectional area),
${\cal C}$ is a universal normalization constant and
we have made use of Eq.~(\ref{bino}) to express the result
in terms of the horizon binormal vectors.

To compare, let us recall that the Wald entropy goes as
\be
S_W\;=\;\oint_{\cal H} s_W dA\;,
\ee
where $dA$ denotes  an area element for a cross-section
of the horizon ${\cal H}$. The density $s_W$ goes (in our notation) as
\be
s_W \;=\; -2\pi\left[{\cal X}^{abcd}\right]^{(0)}\epsilon_{ab}\epsilon_{cd}\;.
\ee
And so the Wald formula
agrees with our expression (\ref{kin-fin3}),
with the normalization now fixed at $\;{\cal C}=-2\pi\;$.

Let us further clarify the relation between the Wald entropy and the
coefficients of the kinetic graviton terms.
We   recall that, for the derivations of the Wald entropy
\cite{WALD,IW,JKM}, the idea was to start with the linearized field equation
\be
\frac{1}{\sqrt{-g}}\left[\frac{\partial\left(\sqrt{-g}{\cal L}\right)}
{\partial g^{ab}}\right]^{(0)}h_{ab}\;=\;0\;
\ee
and reduce this to a boundary term over a cross-section
of the horizon. Following this path, one
ends  up with various terms of
the generic form
\be
\left[{\cal A}^{a_1\ldots a_j;ab}\right]^{(0)}
\bn_{a_1}\ldots \bn_{a_j}h_{ab}
\;. \ee
We now recall
from the previous subsection
that the relevant
gravitons can, when pulled back to the horizon, be exchanged
for  $\;\epsilon_{ab}$'s or  $\bn_a\chi_b\;$'s
(see Eqs.~(\ref{pert},\ref{afterpert})).
Meaning that the Killing identity (\ref{kill}) can be used to reduce
the number of derivatives when
$\;j>0\;$
and, since the Killing vector vanishes on the horizon,
any of these terms can be cast into the form
\be
\left[{\cal A}^{ab}\right]^{(0)} h_{ab}\;.
\label{generic}
\ee

Let us next consider the process of
going from the original volume integral to a surface  integral over
the horizon and, subsequently,  to
a closed  integral over a horizon cross-section. One is then
required to apply Gauss' theorem twice.  Starting with the integrated form of  Eq.~(\ref{generic}) and following this route backwards,
we have
\bea
\oint_{\cal H}\left[{\cal A}^{ab}\right]^{(0)} h_{ab} dA  &\;=\;&
 \int_{\cal H} \left[\partial_{\lambda}\right]^c\bn_c
\left(\left[{\cal A}^{ab}\right]^{(0)} h_{ab}\right)
d\lambda dA \nonumber\\
&\;=\;&  \int_{{\cal M}} \Box\left(\left[{\cal A}^{ab}\right]^{(0)} h_{ab}\right)
\sqrt{-g_{tt}g_{rr}}drdtdA\;,
\eea
with $\lambda$ being the affine parameter for the horizon and ${\cal M}$  the exterior spacetime. Although the first equality is trivial, the second equality is quite complicated, as sensitive
limiting procedures are required
to translate coordinates and  geometric quantities  from the null horizon to
a timelike
``stretched horizon''. The final form also assumes that the graviton
and background do not  depend on the ``non-radial'' spatial coordinates
$x_1,x_2\ldots x_{D-2}$.
Yet, the underlying  message is clear:  Any contribution to the Wald entropy
must necessarily come about from terms in the linearized field equation
(or, equivalently, the quadratic action) carrying two derivatives.
Our analysis explicitly establishes this connection and also
identifies the origin of the kinetic terms.

As observed elsewhere \cite{BGH,BM}, the Wald entropy can be generalized  to other types of  gravitational couplings  by  a different choice of polarization for the gravitons.
For instance, the shear viscosity of a black brane  is determined by the kinetic coefficient
of the $h_{xy}$ gravitons, where $x$ and $y$ are transverse directions on the brane
that are mutually orthogonal as well as orthogonal to the direction of propagation.
On this basis, it had been conjectured \cite{BM} that the
shear viscosity $\eta$ could  be  determined in analogous fashion to
the entropy; that is ({\em cf}, Eq.~(\ref{kin-fin3})),
\be
\eta \;=\; {\cal C}_{\eta} \sum_{a\neq b}^{x,y}\left[{\cal X}^{abcd}\right]^{(0)}
{\tilde \epsilon}_{ab}{\tilde \epsilon}_{cd} \;,
\ee
where ${\tilde \epsilon}_{ab}$ is a suitably defined binormal vector.

However,  our previous use of the Killing vector is unique to the
$h_{rt}$ gravitons on the horizon and, hence, unique to the Wald entropy.  For this reason, the calculation of  any other type of coupling (such as $\eta$) would generally involve the ${\cal Y}$ term in Eq.~(\ref{L-k2}).  This term only becomes relevant for six (or higher) derivative theories. This is because, for a two (four) derivative theory,
$\;{\cal Y}=0 \;$  ($\;{\cal Y}\sim gg\;$); and the would-be kinetic terms  are either identically zero or effectively zero through integration by parts.

\subsection{Summary}

In this section we have  verified our assertion that, for a theory of gravity with any number of derivatives, the kinetic terms
for the $h_{rt}$ gravitons completely account for the Wald entropy.
By expanding out the Lagrangian to second order in
gravitational perturbations,
we have established that, due to the Killing identity, the only  contributing terms are those for which a
single component of the Riemann tensor is responsible for both of the gravitons. This property is essential to the applicability of the Wald formula to higher-derivative gravitational theories.
Indeed, analogue formulas for other  types of graviton coupling
would, as discussed above, break down at the  six-derivative order.

\section{Including matter}

Our considerations have so far  been limited  to  theories of gravity without matter. It is then natural to ask if the inclusion of matter fields could alter any of the  results of  the preceding section.
We now address this question and demonstrate
that, even for a general theory of gravity coupled to matter, all our previous conclusions remain valid.

\subsection{Preliminaries}

Let us now add matter fields, denoted collectively by $\psi$.
Since anti-symmetric combinations of derivatives can always be replaced by
${\cal R}$'s, the Lagrangian is of the form
\be
{\cal L}\;=\;{\cal L}\left[g_{ab},{\cal R}_{abcd},\nabla_{a_1}{\cal R}_{abcd},\nabla_{\left(a_1\right. }\nabla_{\left. a_2\right)}{\cal R}_{abcd},\ldots\;;\;\psi, \nabla_{a_1}\psi,
\nabla_{\left(a_1\right. }\nabla_{\left. a_2\right)}\psi,\ldots\;
\right]\;.
\ee
In principle, one should then add the following set of terms to the expansion of $\;{\cal L}^{(2)}\;$ in Eq.~(\ref{exp}):
\be
\left[\frac{\partial{\cal L}}{\partial\psi}\delta\psi\right]^{(2)}
                         \;+\;\left[\frac{\partial{\cal L}}{\partial[\nabla_{a_1} \psi]}\delta\nabla_{a_1}\psi\right]^{(2)}
 \;+\;\left[\frac{\partial{\cal L}}{\partial[\nabla_{\left(a_1\right.} \nabla_{\left. a_2\right)}\psi]}
\delta\nabla_{\left(a_1\right. }\nabla_{\left. a_2\right)}\psi\right]^{(2)}
\;+\;\ldots\;.
\ee

However,  for scalar (vector, tensor) matter fields, at least the first four (three, two) terms in the series will not contribute.
To understand this point, let us consider
the case of three derivatives
or less acting on a scalar $\phi$. (Two derivatives acting  on a vector and one, on a tensor would be equivalent situations.) As we have seen, a kinetic graviton term can only arise out of  a variation
of the Riemann tensor,
\be
\delta {\cal R}\;\sim\;\nabla\delta\Gamma
+ \delta\Gamma\delta\Gamma\;  \;\;\; {\rm where} \;\;\; \delta\Gamma\sim
\nabla h\;,
\ee
or out of variations of Christoffel symbols.
But recall that  only the $\delta\Gamma\delta\Gamma$ part of the Riemann variation contributes
to the Wald entropy. Hence, a minimal requirement
is having two covariant derivatives that  act as Christoffel symbols.

Now, for a scalar field,   three-derivative terms can be dismissed
because, as  already stressed in Subsection~2.1,
derivatives can only emerge out of  the background in pairs.
The presence of a scalar field (unlike a vector and other odd-spin fields)
cannot viably alter this outcome. This leaves the remaining possibility of
\be
{\cal A}^{ab}\nabla_{\left(a\right.}\nabla_{\left.b\right)}\phi\;.
\ee
At a first glance, this seems to satisfy the minimal requirement.
But, as the right-most derivative is required to act
directly on $\phi$, the above expression reduces to
\be
{\cal A}^{ab}\nabla_{\left(a\right.}\partial_{\left.b\right)}\phi\;,
\ee
and only one Christoffel symbol is available.
It follows that such a term can only induce the variations $\delta g_{ab}$ and
$\delta \phi$; thus disqualifying it as a kinetic contributor.

However, the story could  change
if the matter sector had a sufficiently large
number of
symmetrized  derivatives. The minimum requirements being at least four
derivatives  with a scalar field,
three with a vector or two with a tensor.

\subsection{Four derivative terms}

Let us first consider  the simplest example  of possible contributions:
\be
{\cal L}_{\phi}\; = \;\phi g^{ab} g^{cd}  \nabla_{\left(a\right.}\nabla_b\nabla_c \nabla_{\left.d\right)}\phi \;,
\ee
with $\phi$, again,  a scalar matter field.~\footnote{The  second scalar field
in front is to  prevent this Lagrangian from being a total derivative.}  The variation of
${\cal L}_{\phi}$ with respect to $\;\nabla_{\left(a_{1}\right.}\nabla_{a_{2}}
\nabla_{a_{3}}\nabla_{\left. a_{4}\right)}\phi \;$ will lead to a ``candidate''
 kinetic term of
the form
\be
\left[\delta{\cal L}_{\phi}\right]_k^{(2)} \;=\;\left[\phi g^{ab}g^{cd}\right]^{(0)}\left[\delta \nabla_{\left(a\right.}\nabla_b\nabla_c\nabla_{\left.d\right)}\phi\right]^{(2)} \;.
\ee

A kinetic  term might  appear if any two of the  derivatives act as a Christoffel symbol and the remaining two act directly on the scalar since such a combination  leads  to the the schematic form $\;\left[\delta\left(\Gamma\Gamma \nabla\nabla\phi\right) \right]^{(2)}\;$, and so $\;\Gamma[h] \Gamma [h] \bn\bn\phi \;\sim \bn\bn\phi \bn h\bn h
\;$. Nonetheless,  explicit calculations have indicated that the net kinetic term from
this Lagrangian vanishes.  This finding  can be explained by the
following observations: Even though  ${\cal L}_{\phi}$ appears to have  4!~=~24 distinct terms,
these can be joined into three types
\be
{\cal L}_{\phi} \;=\; 8\phi\Box^2 \phi + 8\phi\nabla^a\nabla^b\nabla_a\nabla_b \phi
+ 8\phi\nabla^a\Box\nabla_a \phi \;.
\ee
One can then use the commutation relations
({\it e.g.},$\;[\nabla_a,\nabla_b] \nabla_c
={\cal R}_{abc}^{\;\;\;\;\;\;\;d}\nabla_d\;$) to iteratively convert the second and third  type into the first.
For instance, one of the 24 terms goes as
\bea
\phi g^{ab}g^{cd}\nabla_a\nabla_c\nabla_b\nabla_d \phi
\;&=&\;\phi g^{ab}g^{cd} \nabla_a\nabla_b\nabla_c\nabla_d \phi \;+\;
\phi g^{ab}g^{cd}  \nabla_a\left[\nabla_c,\nabla_b\right]\nabla_d \phi
\nonumber\\
\;&=&\;\phi \Box^2\phi \;+\;\phi g^{ab}g^{cd}\nabla_a{\cal R}_{cbd}^{\;\;\;\;\;\;e}
\nabla_e\phi\;;\label{woo}
\eea
and similarly for the other 15 terms that are  not  initially of
the $\phi\Box^2\phi$ type.

The end result of the just-described process is
\be
{\cal L}_{\phi} \;=\; 24\phi \Box^2 \phi + \phi
g^{\left(ab\right.}g^{\left.cd\right)} \nabla_{a}\left({\cal R}_{bcd}^{\;\;\;\;\;\;\;e}\
\nabla_e\phi \right)\;,
\label{what}
\ee
with symmetrized  indices on the metrics.
Although there (again) appears to be 4!~=~24 different terms in the Riemann part of the above expression,
this really only contains 16 such terms. The reason
being that 8 of the 24  terms are of the form
\be
\phi g^{ad}g^{bc} \nabla_a {\cal R}_{bcd}^{\;\;\;\;\;\;\;e}
\nabla_e \phi
\;,
\ee
which is already identically vanishing through the contraction of the
first two Riemann indices.

The rest  of the Riemann part vanishes identically as well.
To understand this, let us suppose that the index on the derivative $\nabla_a$
is fixed while the other 3 Riemann indices $bcd$ remain symmetrized.
This leads to the following 6 terms (where we display only the
Riemann tensor for brevity):
\be
 {\cal R}_{\left(bcd\right)}^{\;\;\;\;\;\;\;e} \;=\;
{\cal R}_{bcd}^{\;\;\;\;\;\;\;e} +
  {\cal R}_{dbc}^{\;\;\;\;\;\;\;e} + {\cal R}_{cdb}^{\;\;\;\;\;\;\;e}
+ {\cal R}_{bdc}^{\;\;\;\;\;\;\;e} + {\cal R}_{cbd}^{\;\;\;\;\;\;\;e}
+ {\cal R}_{dcb}^{\;\;\;\;\;\;\;e} \;.
\label{mabel}
\ee
But, the first three terms sum to
zero and, likewise,  the latter three terms,  due to
the Jacobi identity; {\it cf}, Eq.(\ref{jac}).
Then, since  the sum total is (as we vary the index on $\nabla$)
four such vanishing sets, the Riemann part of Eq.~(\ref{what}) is zero.

What is left to establish is that $\;\phi\left[\delta\Box^2 \phi\right]^{(2)}\;$
 similarly makes no  kinetic contribution.
This can, indeed, be verified with an explicit
calculation but also follows from a simple argument.
To show this, let us first recall that
\be
\Box^2 \phi \;=\; \frac {1}{\sqrt{-g}}\partial_a \left\{ g^{ab} \sqrt{-g} \partial_b\left[ \frac{1}{\sqrt{-g}}
\partial_c\left( g^{cd} \sqrt{-g} \partial_d \phi \right)\right] \right\}\;.
\ee
After we disregard all  kinetic contributions
that can be gauged away, which include either derivatives acting on determinants or derivatives acting transversely, one finds that the  sole potential contributor is
\be
\Box^2\phi \;\to\;
g^{ab}\left[\partial_a \partial_b g^{cd}\right]\left[\partial_c\partial_d \phi
\right]\;.
\ee
However, even this term can still be gauged away  through integration
by parts, and so vanishes.

\subsection{2n derivative terms}

The same trend extends to any  number of symmetrized
derivatives. Again working with a scalar matter field $\phi$ for simplicity,
we can justify this claim via the following arguments:

Let us begin by considering the Lagrangian
\be
{\cal L}_{2n}(\phi)\;=\; \phi g^{a_1a_2}\cdots g^{a_{2n-1}a_{2n}}\nabla_{\left(a_1\right.}
\nabla_{a_2}\dots\nabla_{a_{2n-1}}\nabla_{\left. a_{2n}\right)}\phi \;,
\ee
which contains $2n$ symmetrized derivatives.
Similarly to  the four-derivative case, ${\cal L}_{2n}(\phi)$ should decompose into the form
\be
\phi\Box^n \phi\;\;\; + \;\;\;
\;\phi\sum^{n-1}_{k=1}\left[{\cal R}^{[k]}\nabla^{[2n-2k-1]}\right]^a \nabla_a\phi\;,
\ee
with the square bracket meant to  represent a collection of $k$ (4-index) Riemann tensors
followed by $2n-2k-1$ $\nabla$'s contracted in all possible ways.  Such an arrangement
is always possible given a sufficient number of a  commutations
of derivatives.~\footnote{To ensure the displayed ordering, one should commute derivatives
from left to right; {\it i.e.}, opposite to the direction of Eq.~(\ref{woo}).}

Now, the key point is
that any of the generated Riemann tensors arises
due to a commutation of symmetrized $\nabla$'s, and
so is of the basic form
\be
\left[\nabla_{\left(a_i\right.},\nabla_{a_j}\right]\nabla_{\left.a_k\right)}\;\dots \;=\;
{\cal R}_{\left(a_ia_ja_k\right)}^{\;\;\;\;\;\;\;\;\;\;\;\;\;e}\;\nabla_e\;\dots\;,
\ee
whereby  three of the Riemann indices are explicitly symmetrized while the
fourth index is summed over independently.
So that, just like for the $4$-derivative example (Eq.~(\ref{mabel})) any Riemann
tensor produces  two sets of 3 terms with each vanishing
due to the Jacobi identity.

Let us now focus on the $\;\phi\Box^n\phi\;$ term.
Again disregarding kinetic contributions that
can be gauged away,  we are left with potential contributors of only the two
 basic forms,
\be
\dots g^{ab}g^{cd}\partial_a h^{ef} \partial_{c} h^{ij} \dots \partial_b
\partial_d \partial_e \partial_f \partial_i\partial_j \phi
\;\;\; {\rm and}
\;\;\; \dots g^{ab}\partial_a h^{cd} \partial_{b} h^{ij}\dots\partial_c\partial_d
\partial_i\partial_j \phi
 \;,
\ee
with the dots indicating other irrelevant structure.
The crucial point here is that
the derivatives act only symmetrically on the gravitons,
whereas a Riemann tensor is constructed out of  anti-symmetric
combinations of derivatives.
Meaning that these terms
are simply  not capable of hiding a Riemann tensor.

\subsection{Summary}

Given a generalized theory of gravity, we have now shown that
adding  matter fields with any number of symmetrized covariant derivatives
acting on them
does not lead to any  kinetic contributions beyond those
already encountered in Section~3. The explicit calculations used scalar fields
but  can, as discussed, be generalized in a straightforward manner to any type of tensor field.

\section{The generalized field equation}

The calculation of the quadratic action for a generalized theory is often complicated.
But one can rather use  the linearized field equation
as an equivalent but simpler means for extracting the Wald entropy.
The compatibility of the two approaches follows from their
equivalency up to total derivatives.
An appropriate choice of surface term can always be used
to cancel a total derivative and then,
as explained in \cite{JKM}, any such boundary term
does not contribute to the Wald formula.

The entropy should be extracted from terms of the schematic form
$\nabla \nabla h $ in the field equation, subject to  the various subtleties discussed. The field equation has already
been presented (without a full derivation) in \cite{IW}.
However, for practical calculations, we have identified a
form of the equation differing  by a sign from that of \cite{IW}.

\subsection{The linearized field equation}

Let us presume, for simplicity in the presentation, that
the matter fields $\psi$ are coupled only to the metric and any such terms have been collected separately in $\frac{1}{2}{\cal L}_{\psi}$. Then the gravitational field equation for the density
\be
{\cal L}_{total}\;=\;\sqrt{-g}\left[{\cal L}+\frac{1}{2}{\cal L}_{\psi}\right]\;
\ee
is
\be
 \frac{\partial{\cal L}}{\partial g_{pq}}\delta g_{pq}
\;+\; \frac{1}{2}g^{pq}{\cal L}\delta g_{pq}
\;+\;{\cal X}^{abcd}\delta{\cal R}_{abcd}[\delta g_{pq}]
\;+\;{\cal Q}^{pq}\delta g_{pq}
\;=\; -T^{pq}\delta g_{pq}\;,
\label{linear2}
\ee
where
\be
T^{pq}\;\equiv\;
\frac{2}{\sqrt g}\frac{\partial\left[\sqrt{-g}{\cal L}_{\psi}\right]}
{\partial g_{pq}}\;,
\ee
${\cal X}^{abcd}$ is defined in Eq.~(\ref{W2}) and the tensor
${\cal Q}^{pq}$ accounts for
the extraneous terms in the iterative procedure of Section 2.
(But, if ${\cal L}$  contains only
anti-symmetrized combinations of derivatives, there would be no
${\cal Q}$ contribution.)
Note that, had we varied with respect to the contravariant form of
the metric $\delta g^{pq}$, then the stress tensor needs to be defined as
\be
T_{pq}\;\equiv \;-\frac{2}{\sqrt g}
\frac{\partial\left[\sqrt{-g}{\cal L}_{\psi}\right]}
{\partial g^{pq}}\;.
\ee

Now linearizing and rearranging, we have
\be
\left[{\cal X}^{abcd}\right]^{(0)}\delta{\cal R}^{(1)}_{abcd}[h_{pq}]
\;=\; -\left[T^{pq}\;+\;\frac{\partial{\cal L}}{\partial g_{pq}}
\;+\; \frac{1}{2}g^{pq}{\cal L} \;+\;{\cal Q}^{pq}\right]^{(0)} h_{pq}\;,
\label{linear}
\ee
with the right-hand side being irrelevant to
the Wald entropy.

We will thus focus on the left-hand side and recall the expansion for $\delta {\cal R}_{abcd}^{(1)}$ in
Eq.~(\ref{expan}).
This expression and the symmetry properties of the
background  Riemann tensor
(which are shared by $\left[{\cal X}^{abcd}\right]^{(0)}$)
indicate that the left-hand side
of Eq.~(\ref{linear}) reduces to the sum of four equivalent terms.
Denoting this sum as
${\cal G}$, we obtain
\bea
{\cal G}\;=\;\left[{\cal X}^{abcd}\right]^{(0)}\delta{\cal R}^{(1)}_{abcd}[h_{pq}]
\;&=&\; 2 \left[{\cal X}^{apqb}\right]^{(0)}\bn_a\bn_b h_{pq} \nonumber\\
\;&=&\; {\cal G}^{pq}h_{pq}\;.
\label{E2}
\eea
For the rest of this section, it is implied that the
zeroth-order geometry applies to all tensors
besides the $h$'s.

To proceed, one considers separately the symmetric
$\frac{1}{2}\{\nabla_a,\nabla_b\}$ and the anti-symmetric
$\frac{1}{2}[\nabla_a,\nabla_b]$ combinations of the derivatives.
Whereas the former is trivially handled with a double
integration by parts, the latter is  more complicated.
Nonetheless, it is possible to convert $\{\nabla_a,\nabla_b\}$ into
Riemann tensors. Then, through repeated application of the
symmetries of both the ${\cal X}$ and ${\cal R}$ tensors, one obtains the expression in the second line of Eq.~(\ref{E2}). What is left is
to strip off the graviton and then linearize ${\cal G}_{pq}$.

Skipping over subtleties, we find that  the complete field equation   now  goes as ({\it cf},
Eq.~(\ref{linear}))
\be
2\nabla_b\nabla_a {\cal X}^{apqb} \;-\;
{\cal X}^{abcp}{\cal R}_{abc}^{\;\;\;\;\;\;\;q}
\; +\; \frac{1}{2}g^{pq}{\cal L}
\;=\;
-T^{pq}\;,
\label{fieldeq}
\ee
where we have assumed that ${\cal L}={\cal L}\left[{\cal R}_{abcd}\right]$
for simplicity. This is in contrast to the relative positive sign between the first terms in Eq.~(\ref{fieldeq}) in \cite{IW}.

\subsection{Simple examples}

To clarify  this process, let us recall
the Einstein Lagrangian, $\;{\cal L}_{Ein}={\cal R}\;$, regarded as independent of the metric.
Plugging this into  Eq.~(\ref{W2}), we  find that
\be
{\cal X}_{Ein}^{abcd} \;=\; \frac{1}{2}\left[g^{ac}g^{bd}-g^{ad}g^{bc}\right]\;,
\ee
and the tensor ${\cal G}^{pq}$ of  Eq.~(\ref{E2}) then  reduces to
\be
{\cal G}_{Ein}^{pq} = -{\cal R}^{pq} \;.
\ee
This is indeed the correct form, as substituting back into
the field equation (\ref{fieldeq}), we obtain
\be
{\cal R}^{pq} \;-\frac{1}{2}g^{pq}{\cal R} \;=\; T^{pq}\;.
\ee

Let us further confirm the consistency of  Eq.~(\ref{E2})
by starting with an $\;{\cal L}=F({\cal R})\;$ theory
of gravity ($\;{\cal R}=g^{ac}g^{bd}{\cal R}_{abcd}\;$) and
following the standard procedure,
\be
 \partial_{\cal R}F({\cal R}) \left[{\cal R}_{pq}\delta g^{pq}
\;+\;g^{pq}\delta {\cal R}_{pq}\right] \;-\;\frac{1}{2}
F({\cal R})g_{pq}\delta g^{pq}
 \;=\; T_{pq} \delta g^{pq} \;.
\ee

To handle the Ricci variation, one can use a contracted form
of  Eq.~(\ref{expan}) and then integrate by parts (twice) to
free up the graviton. In this way, one ultimately finds that
({\it e.g.}, \cite{f-of-R})
\be
\left[{\cal R}_{pq} + g_{pq}\Box - \nabla_p\nabla_q  \right]\partial_{\cal R}
 F({\cal R})\;-\; \frac{1}{2}g_{pq}{\cal F}({\cal R})
\;=\; +T_{pq} \;
\ee
or, equivalently,
\be
\left[\nabla^p\nabla^q - g^{pq} \Box - {\cal R}^{pq}    \right]\partial_{\cal R}
 F({\cal R})\;+\; \frac{1}{2}g^{pq}{\cal F}({\cal R})
\;=\; -T^{pq} \;.
\ee

One can verify that this is in perfect agreement
with Eq.~(\ref{fieldeq}), given that
\bea
{\cal X}^{apqb}\;&=&\; \frac{\partial F({\cal R})}{\partial {\cal R}_{apqb}}
\nonumber \\
\;&=&\;\frac{1}{2}\left[g^{aq}g^{pb}-g^{ab}g^{pq}\right]
\partial_{\cal R}F({\cal R})
\;.
\eea

\section{Example calculations}

\subsection{General considerations}

We wish to demonstrate by specific examples that the  coefficients
of the $h_{rt}$ kinetic terms can indeed be used to
directly extract the Wald entropy. We will use two gravitational theories,
\be
{\cal L}_{\alpha}\;=\;
{\cal R} \;+\;
\alpha
{\cal R}^{abcd}{\cal R}_{abcd}\;,
\label{alpha}
\ee
\be
{\cal L}_{\beta}\;=\;
{\cal R} \;+\;
\beta\nabla^{k}R^{abcd}\nabla_{k}R_{abcd}\;,
\label{beta}
\ee
where $\alpha$ and $\beta$ are
constants.  From a physically motivated perspective, these parameters
should be regarded as small:
$\;\alpha
\ll r_h^2 \;$ and $\;\beta \ll r_h^4\;$,
where $r_h$ is the horizon radius of the corresponding black hole solution.

The Einstein term ${\cal R}$ allows us to
normalize our results and, more importantly, given that the corrections to Einstein gravity are small,
it allows us to use the Einstein background solution in the
calculations. Any contribution from a perturbative correction is already first order in $\alpha$ or $\beta$, so that the Einstein background solution suffices. The kinetic coefficients in Einstein's theory  are numerical constants and so are insensitive to the form of the solution.~\footnote{Here,
it is assumed that the intention is to calculate
the Wald entropy in units of horizon area. The Einstein contribution
to the horizon area does get corrected.}

In the following, we incorporate the notation
$\;\overline{g}_{ab}=g_{ab}^{(0)}\;$, $\;{\overline{\cal R}}_{abcd} = {\cal R}^{(0)}_{abcd}\;$
and
$\;\overline{\Gamma}_{ab,c}=\Gamma^{(0)}_{ab,c}\;$, where
$\;\Gamma_{ab,c}=g_{cd}\Gamma^{d}_{ab}\;$.

\subsection{4-derivative gravity}

Let us first  consider the theory Eq.~(\ref{alpha}).
Keeping in mind that the Lagrangian  is integrated to get an action,
$I=\int\sqrt{-g}{\cal L}_{\alpha}d^{D}x$,
we can integrate by parts.

To begin, we  write the linear in $\alpha$ and quadratic in $h$ part of the Lagrangian density
\be
\delta\left[{\sqrt g}{\cal L}_{\alpha}\right]^{(2)}\;=\;
\alpha\delta\left[\sqrt{-g}g^{a\alpha}g^{b\beta}g^{c\gamma}g^{d\delta}{\cal R}_{\alpha\beta\gamma\delta}{\cal R}_{abcd}\right]^{(2)}\;
\label{xxx1}
\ee
and  look for any term that makes a kinetic contribution. As a first step, let us consider the ``(1,1)'' terms
or
$\delta\left[\sqrt{-g}g^{a\alpha}g^{b\beta}g^{c\gamma}g^{d\delta}{\cal R}_{\alpha\beta\gamma\delta}\right]^{(1)}
\delta{\cal R}_{abcd}^{(1)}$. Our previous analysis suggests that it does not contribute. We will demonstrate this
explicitly by  working with the  unexpanded version
of the factor $\sqrt{-g}g^{a\alpha}g^{b\beta}g^{c\gamma}g^{d\delta}{\cal R}_{\alpha\beta\gamma\delta}$.
So we rewrite Eq.~(\ref{xxx1}) as
\begin{equation}
\delta\left[{\sqrt g}{\cal L}_{\alpha}\right]^{(2)}\;=\;
=\alpha\sqrt{-g}g^{a\alpha}g^{b\beta}g^{c\gamma}g^{d\delta}{\cal R}_{\alpha\beta\gamma\delta} \left[\bn_{b}\bn_{c}h_{ad}-\bn_{b}\bn_{d}h_{ac}- \bn_{a}\bn_{c}h_{bd}+\bn_{a}\bn_{d}h_{bc}\right]\;,
\end{equation}
where  Eq.~(\ref{expan}) has been used to expand out $\delta{\cal R}_{abcd}^{(1)}$.

Now, to order $h$ in the square brackets,
any of the $\bn$'s can be replaced  with a ``full'' $\nabla$.
Let us do so with the left-most $\bn$ in each of the four terms
and then integrate by parts:
\begin{equation}
-\alpha\sqrt{-g}g^{a\alpha}g^{b\beta}g^{c\gamma}g^{d\delta}\left[\nabla_{b}{\cal R}_{\alpha\beta\gamma\delta}\bn_{c}h_{ad}-\nabla_{b}{\cal R}_{\alpha\beta\gamma\delta}\bn_{d}h_{ac}-\nabla_{a}{\cal R}_{\alpha\beta\gamma\delta}\bn_{c}h_{bd}+\nabla_{a}{\cal R}_{\alpha\beta\gamma\delta}\bn_{d}h_{bc}\right]\;.
\label{xxxxx2}
\end{equation}

As in Section 2, when two  $\nabla$'s act on a graviton, the Killing relation can be used to convert them to a Riemann tensor. So, to get a kinetic term out of Eq.~(\ref{xxxxx2}), we need to expand ${\cal R}_{\alpha\beta\gamma\delta}$ to  first order in $h$ and to zeroth order in derivatives. Then, since
$\;{\cal R}_{abcd}=\partial_{b}\Gamma_{ac,d}- \partial_{a}\Gamma_{bc,d}+g^{mn}\Gamma_{ac,m} \Gamma_{nb,d}-g^{mn}\Gamma_{bc,m}\Gamma_{na,d}\;$
and $\;\Gamma_{ab,c}=\frac{1}{2}\left(\partial_{a}g_{bc}+\partial_{b}g_{ac}-\partial_{c}g_{ab}\right)\;$,
the only terms in ${\cal R}_{abcd}$ that can possibly contribute
are $g^{mn}\overline{\Gamma}_{ac,m}\overline{\Gamma}_{nb,d}$
and   $g^{mn}\overline{\Gamma}_{bc,m}\overline{\Gamma}_{na,d}\;$.
But  $\;\nabla_b g^{mn} = \nabla_a g^{mn} =0\;$, so  there can
be no kinetic contribution at all.

Hence, we have
\be
\delta\left[{\sqrt g}{\cal L}_{\alpha}\right]^{(2)}\;=\;
2\alpha\left[\sqrt{-g}{\cal R}^{abcd}\right]^{(0)}\delta{\cal R}_{abcd}^{(2)}+\cdots\;,
\ee
or,  using Eq.~(\ref{rtoo}),
\be
\delta\left[{\sqrt g}{\cal L}_{\alpha}\right]^{(2)}\;=\;
2\alpha\sqrt{-{\overline g}}{\overline{\cal R}}^{abcd}\left[\bn_{m}h_{bc}\bn^{m}h_{ad}+2\bn^{m}h_{ac}\bn_{b}h_{dm}\right]\;,
\ee
where the second term in the square brackets can be gauged away.

We can identify $\;s_W\;=\;2\alpha{\cal C}{\overline{\cal R}}^{abcd}\epsilon_{ab}\epsilon_{cd}\;$,
with the normalization $\;{\cal C}=-2\pi\;$  given by the Einstein term. This is in agreement with that obtained via a direct application of Wald's formula.

\subsection{6-derivative gravity}

The Lagrangian is now given by Eq.~(\ref{beta}).
Here, we begin the quadratic density
\be
\delta\left[{\sqrt g}{\cal L}_{\alpha}\right]^{(2)}\;=\;
\beta\delta \left[\sqrt{-g}g^{a\alpha}g^{b\beta} g^{c\gamma}g^{d\delta}\nabla_k{\cal R}_{\alpha\beta\gamma\delta}\nabla^k{\cal R}_{abcd}\right]^{(2)}\;.
\label{xxx2}
\ee
The (1,1) terms, again, cannot contribute.
Because of the derivatives in front of the Riemann tensors, either of
these must be expanded to first order in $h$ and zeroth order in derivatives. Hence, by the very same reasoning as provided above, a kinetic contribution can not be obtained.

This leaves us, after integration by parts, with
\be
\delta\left[{\sqrt g}{\cal L}_{\beta}\right]^{(2)}\;=\;
-2\beta\left[\sqrt{-g}\nabla^k\nabla_k{\cal R}^{abcd}\right]^{(0)}\delta{\cal R}_{abcd}^{(2)}+\cdots\;,
\ee
or, like before,
\be
\delta\left[{\sqrt g}{\cal L}_{\alpha}\right]^{(2)}\;=\;
-2\beta\sqrt{-{\overline g}}\bn^k\bn_k
\overline{{\cal R}}^{abcd}\left[\bn_{m}h_{bc}\bn^{m}h_{ad}+ 2\bn^{m}h_{ac}\bn_{b}h_{dm}\right]+\cdots\;,
\ee
and so the identification
$\;s_W\;=\;-2\beta{\cal C}\Box{\overline{\cal R}}^{abcd}\epsilon_{ab}\epsilon_{cd}\;$
follows. This is, once again, in agreement with the Wald formula when the normalization is $\;{\cal C}=-2\pi\;$.

Finally,  more derivatives
beyond six would neither conceptually nor technically
complicate the calculation.


\section{Conclusion}

To summarize, we have investigated Wald's  Noether charge entropy
\cite{WALD,IW}, relying on its identification with a quarter of the horizon area in units of the effective gravitational coupling, as first established in \cite{BGH}. The Wald entropy can, as now verified, be  determined from
the kinetic coefficients
for the $h_{rt}$ gravitons on the horizon.
We have also clarified what terms in the
quadratically expanded action
(or linearized field equation) can contribute to the entropy and illustrated our general procedure with
some of  explicit examples. Additionally, we  have reconsidered
the gravitational field equation for a general  theory of gravity
and  presented it in a form which differs from \cite{IW}.

\section{Acknowledgments}

The research of RB was supported by The
Israel Science Foundation grant no 239/10.
The research of MH  was supported by The Open University of Israel's Research Fund and by Israel Science Foundation grant no 239/10.
The research of AJMM was supported by the Korea
Institute for Advanced Study and Rhodes University.
AJMM thanks  Ben-Gurion University for their hospitality during his visit. We thank Aaron Amsel for useful discussions

\end{document}